# A Glimpse of the First Galaxies

Naveen A. Reddy (National Optical Astronomy Observatory)

*The recently refurbished Hubble Space Telescope reveals a galaxy from a time when the Universe was just 500 million years old, providing insights into the first throes of galaxy formation and the reionization of the Universe.*

A central focus of cosmology is to understand how the primordial density fluctuations imprinted by the Big Bang gave rise to the galaxies and larger structures we observe today. Just as archaeologists sift through deeper layers of sand to uncover the past, cosmologists use large telescopes and sensitive detectors to study galaxies at ever greater distances from Earth and, because of the finite speed of light, to peer farther back in time. On page 504 of this issue, Bouwens *et al.*[1] take another step in this direction by exploiting the deepest near-infrared images of the sky, which were obtained with the re-serviced Hubble Space Telescope and its new Wide Field Camera[2]. On the basis of these data, the authors report the plausible detection of the most distant galaxy yet discovered. The galaxy would have existed when the Universe was just 4% of its current age and when one of the most important phase transitions of gas in the Universe occurred.

Building on previous studies, Bouwens and colleagues used the well-established Lyman break technique[3] to select galaxies at the largest distances, or redshifts. The method relies on the absorption, by neutral hydrogen within a galaxy or by intervening hydrogen clouds, of photons that are more energetic than Lyman-$\alpha$ photons (10.2 eV, corresponding to a wavelength of 1,216 ångströms). The resulting decrease in flux bluewards of the Lyman-$\alpha$ wavelength results in a characteristic 'break' in the spectrum of a galaxy. Galaxies at different redshifts can then be located by searching for objects that are detected in one filter but that disappear, or are very faint, in a bluer filter.

Until now, the primary obstacle to identifying galaxies beyond redshift 6 — when the Universe was less than 1 billion years old — has been that the Lyman break shifts to the observed near-infrared, where the emission from the sky background is several hundred times higher than it is in the visible range of the spectrum. This higher background inhibits the ability to obtain deep imaging, and has motivated observations from above Earth's atmosphere. A breakthrough came with the installation of the Wide Field Camera on Hubble; the camera's increased field of view and sensitivity over the previous near-infrared instrument on Hubble results in an increase by a factor of more than 30 in its capacity for finding faint galaxies at high redshift.

Using multi-filter imaging from Hubble and the Lyman break technique, Bouwens and collaborators[1] report the discovery of one candidate galaxy at redshift of about 10 (Fig. 1). Comparing the number density of galaxies at redshift 10, inferred from their observations, with that determined at lower redshifts, they find that the average galaxy increases in luminosity by more than a factor of 10 during the first 2 billion years of galaxy formation. Taken one step further, this finding suggests a close connection between galaxy formation and the assembly of dark matter in the early Universe.

In contrast to the prevailing theory of cold dark matter and its relative success in reproducing the large-scale structure of the Universe, the physics of the development and evolution of visible matter is difficult to model: it depends on complex processes that govern the cooling of gas to form stars, the evolution of the stars themselves, and the feedback of energy and matter from stars and black holes. It is perhaps remarkable, therefore, that at early cosmic times the growth of galaxies seems to mirror that of the dark-matter halos in which the galaxies reside[4]. This similarity suggests that, despite the seemingly complex physics of star formation, simple gravitational theory — combined

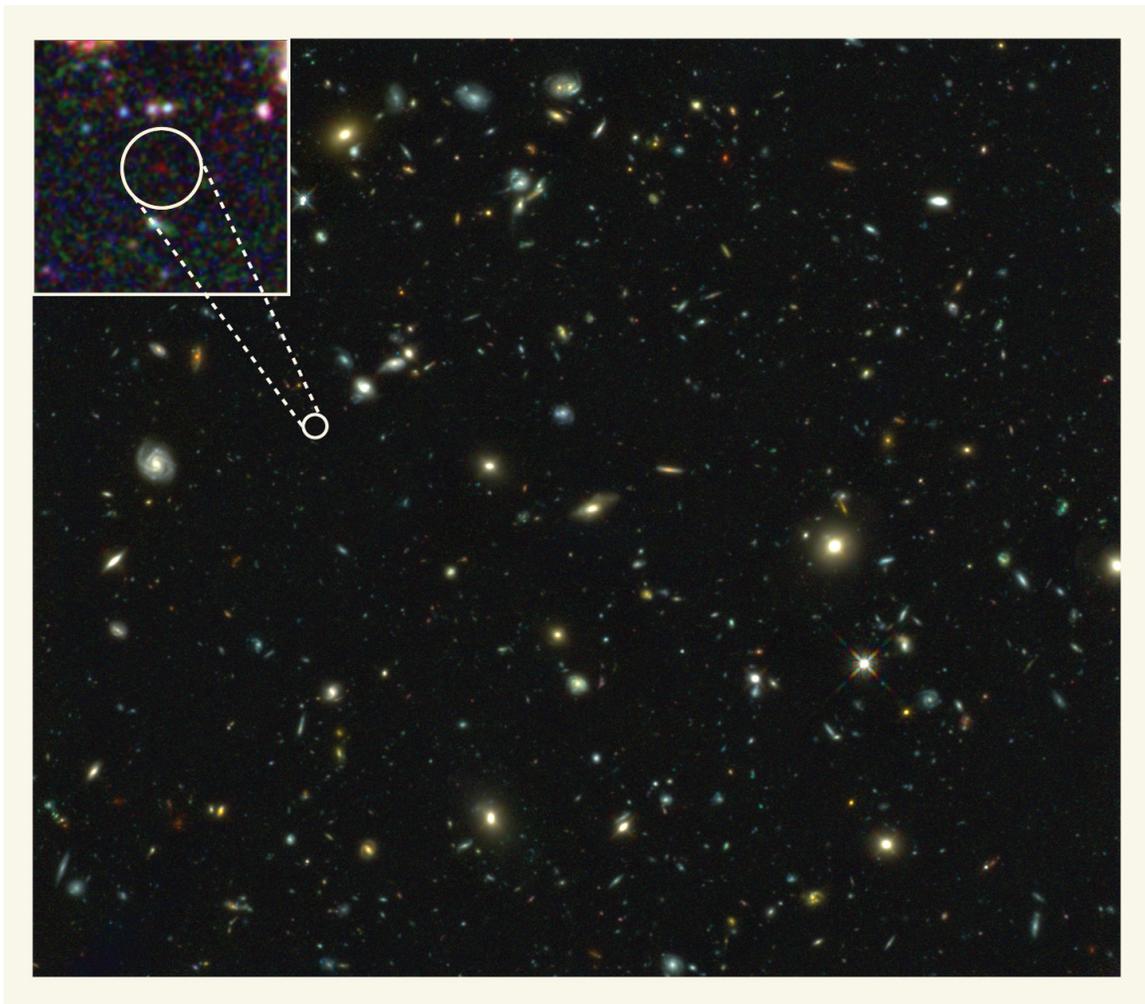

**Figure 1: A galaxy at redshift 10.** Bouwens and colleagues' search[1] for galaxies in the Hubble Ultra Deep Field has resulted in the plausible detection of the most distant galaxy yet detected. The circle marks the location of the galaxy (red blob in inset). [Credit- NASA/ESA/G. Illingworth (UCO/Lick Obs. & Univ. California, Santa Cruz)/R. Bouwens (UCO/Lick Obs. & Leiden Univ.)/HUDF09 TEAM]

with a factor that parameterizes the efficiency of star formation (or, the fraction of gas that is converted to stars) — can provide a first-order prediction of the luminosity of a galaxy.

Aside from probing the earliest stages of galaxy formation, a topical area of interest in cosmology is to identify the sources responsible for the transition between a neutral state of hydrogen in the Universe (roughly 300,000 years after the Big Bang) to a mostly ionized state at redshift about 6 (950 million years after the Big Bang). Bouwens and colleagues' study[1] probes galaxies at the heart of this 'reionization' epoch. Given some — albeit very uncertain — assumptions of the clumpiness of gas in the Universe and the fraction of ionizing photons that can escape galaxies, they argue that galaxies at redshift 10 may not provide enough ultraviolet flux to reionize the Universe. The dominant contributor to the ionizing flux at early cosmic epochs remains a mystery. Nonetheless, the plausible detection of a galaxy at redshift 10 suggests an onset of star formation at redshift beyond 12 (about 100 million years earlier), potentially increasing the role of galaxies in the early ionization of the Universe.

Although these results[1] give us a glimpse of the earliest stages of galaxy formation, substantial uncertainties remain and more work is needed. Sample variance remains the dominant uncertainty, as a result of the small number of objects and the small field of view surveyed (equivalent to an area of about 0.6% the size of the Moon). Even more crucial, however, is the need to confirm the redshifts of these objects. The best confirmation of distance would be the detection of a strong emission line in the spectrum, such as the Lyman-$\alpha$ line. Detecting this line may be challenging for these 'primordial' galaxies because they are expected to be gas-rich (having not had enough time to convert a significant fraction of their gas into stars) and to be surrounded by a mostly neutral medium that resonantly scatters Lyman-$\alpha$ photons.

The best hope is the James Webb Space Telescope (JWST). With its larger mirror and near-infrared-sensitive detectors, this facility will dramatically improve the situation: imaging and spectroscopy across a larger swathe of the spectrum will enable the confirmation of a spectral break or the detection of a strong emission line. Scheduled for launch in 2014, the JWST will also have the sensitivity to detect galaxies at redshift 10 that are even fainter than the one reported by Bouwens and collaborators. Studying this faint population will yield a more complete picture of their role in reionizing the Universe. The authors' preliminary foray in studying the first galaxies underscores the important role of facilities such as the JWST in revolutionizing our understanding of galaxy formation at the earliest cosmic epochs, and paves the way for a bright future in studying faint and distant galaxies.

Naveen A. Reddy is at the National Optical Astronomy Observatory, Tucson, Arizona 85719, USA.
e-mail: nar@noao.edu